\documentclass[preprint]{aastex7}

\usepackage{enumerate}

\usepackage{tabularx}
\usepackage{array}
\usepackage{verbatim}
\usepackage{graphicx}
\usepackage{ulem}
\usepackage{hyperref}

\begin{document}

\title{Identifying eclipsing binary stars with TESS data based on a new hybrid deep learning model}

\author{Ying Shan}\email{1222045516@njupt.edu.cn}
\affiliation{College of Computer, Nanjing University of Posts and Telecommunications, Nanjing 210023, China}
\affiliation{Jiangsu Key Laboratory of Big Data Security and Intelligent Processing, Nanjing 210023, China}

\author[0000-0001-8869-653X]{Jing Chen}\email{chenjing@niaot.ac.cn}
\affiliation{Nanjing Institute of Astronomical and Optics \& Technology, Chinese Academy of Sciences, Nanjing 210042, China
}
\affiliation{CAS Key Laboratory of Astronomical Optics \& Technology, Nanjing Institute of Astronomical Optics \& Technology, Nanjing 210042, China}

\author[0009-0006-0874-3273]{Zichong Zhang}\email{zczhang2022@niaot.ac.cn}
\affiliation{Nanjing Institute of Astronomical and Optics \& Technology, Chinese Academy of Sciences, Nanjing 210042, China
}
\affiliation{CAS Key Laboratory of Astronomical Optics \& Technology, Nanjing Institute of Astronomical Optics \& Technology, Nanjing 210042, China}
\affiliation{University of Chinese Academy of Sciences, Beijing 100049, China}

\author[0000-0003-3603-1901]{Liang Wang}\email{liangwang@niaot.ac.cn}
\affiliation{Nanjing Institute of Astronomical and Optics \& Technology, Chinese Academy of Sciences, Nanjing 210042, China
}
\affiliation{CAS Key Laboratory of Astronomical Optics \& Technology, Nanjing Institute of Astronomical Optics \& Technology, Nanjing 210042, China}
\affiliation{University of Chinese Academy of Sciences, Beijing 100049, China}

\author[0000-0003-2828-8491]{Zhiqiang Zou}\email{zouzq@njupt.edu.cn}
\affiliation{College of Computer, Nanjing University of Posts and Telecommunications, Nanjing 210023, China}
\affiliation{Jiangsu Key Laboratory of Big Data Security and Intelligent Processing, Nanjing 210023, China}
\affiliation{University of Chinese Academy of Sciences, Nanjing, Jiangsu, 211135, People’s Republic of China}

\author{Min Li}\email{zouzq@njupt.edu.cn}
\affiliation{College of Computer, Nanjing University of Posts and Telecommunications, Nanjing 210023, China}
\affiliation{Jiangsu Key Laboratory of Big Data Security and Intelligent Processing, Nanjing 210023, China}

\correspondingauthor{Liang Wang}
\email{liangwang@niaot.ac.cn}
\correspondingauthor{Zhiqiang Zou}
\email{zouzq@njupt.edu.cn}

\begin{abstract}
Eclipsing binary systems (EBs), as foundational objects in stellar astrophysics, have garnered significant attention in recent years. These systems exhibit periodic decreases in light intensity when one star obscures the other from the observer’s perspective, producing characteristic light curves (LCs). With the advent of the Transiting Exoplanet Survey Satellite (TESS), a vast repository of stellar LCs has become available, offering unprecedented opportunities for discovering new EBs. To efficiently identify such systems, we propose a novel method that combines LC data and generalized Lomb-Scargle periodograms (GLS) data to classify EBs. At the core of this method is CNN\_Attention\_LSTM Net (CALNet), a hybrid deep learning model integrating Convolutional Neural Networks (CNNs), Long Short-Term Memory (LSTM) networks, and an Attention Mechanism based on the Convolutional Block Attention Module (CBAM). We collected 4,225 EB samples, utilizing their 2-minute cadence LCs for model training and validation. CALNet achieved a recall rate of 99.1\%, demonstrating its robustness and effectiveness. Applying it to TESS 2-minute LCs from Sectors 1 to 74, we identified 9,351 new EBs after manual visual inspection, significantly expanding the known sample size. This work highlights the potential of advanced deep-learning techniques in large-scale astronomical surveys and provides a valuable resource for further studies on EBs.

\end{abstract}


\section{Introduction} \label{sec1}
Eclipsing binary systems (EBs) are binary star systems in which the components periodically pass in front of each other, resulting in periodic dimming in their light curves (LCs). The periods, shapes, durations, and depths of these LCs reveal the fundamental parameters of the binary systems, such as the stellar radii, temperatures, luminosities, orbital inclinations, and semi-major axes of the binaries. Some EBs are also spectroscopic binaries, allowing the measurement of radial velocity variations over an orbital period. Combining radial velocities with LCs makes it possible to determine the masses and densities of the stars precisely. This method is independent of stellar evolution models, making EBs good laboratories in stellar astrophysics for testing theories of stellar evolution \citep{feiden2012reevaluating}, distances \citep{stassun2016eclipsing}, asteroseismic theories \citep{gaulme2016testing}, and gravitational theories \citep{baroch2021analysis}.

Over the past decade, ground-based and space-based photometric surveys have significantly advanced the discovery of new EBs and opened a new window into these systems. For example, \cite{papageorgiou2018updated} presented a catalog containing 4680 EBs in the northern sky using data from Catalina Sky Surveys (CSS). \cite{chen2020zwicky} present a catalog of more than 781,000 periodic variables from  Data Release 2 of Zwicky Transient Facility (ZTF), of which 350,000 were classified as EBs. The space-based Kepler mission \citep{borucki2016kepler} nearly continuously observed $\sim2$ million stars within a sky field of 105\,deg$^2$ and found over 2,800 eclipsing and ellipsoidal binary systems \citep{kirk2016kepler}. As the successor of Kepler, the Transiting Exoplanet Survey Satellite (TESS) is scanning almost the entire sky with a much larger field of view to search for extra-solar planets \citep{ricker2015transiting}, and its data have been proven to be invaluable for EB search. \cite{prvsa2022tess} identified 4,584 EBs using the first two years of 2-min cadence data (Sectors 1--26). Furthermore, \cite{ijspeert2021all} analyzed the 30-min cadence data in the same sectors and found 3155 OBA-type EB candidates. These works highlight the power of finding new EBs using TESS data.

In the analysis of photometric survey data, the huge volume of data poses significant challenges for the identification of new objects, based on the results of binary classification (EBs and non-EBs). Previous studies have demonstrated that machine learning (ML) techniques have great potential for classifying and detecting weak signals in large photometric datasets. For example, \cite{papageorgiou2019physical} used the artificial neural network (ANN) to analyze the physical properties of over 2200 EBs identified by the Catalina Sky Survey. \cite{shallue2018identifying} presented a machine-learning approach for identifying new transiting exoplanets in the Kepler data. \cite{osborn2020rapid} applied convolutional neural networks (CNNs) to classify the transit signals detected by TESS, enhancing the efficiency of candidate validation. Additionally, \cite{ofman2022automated} developed a new ML method applied to threshold crossing events (TCEs) in the exoplanet transit signal search, resulting in three new exoplanet candidates.

In our study, we collected LC data from TESS and incorporated generalized Lomb-Scargle periodograms (GLS) \citep{zechmeister2009generalised} to capture periodic features from LC. By combining these two types of data, we developed a novel deep learning method for identifying EBs.

The organization of this paper is as follows: Section \ref{sec2} introduces the two types of stellar datasets utilized in this study, along with the preprocessing method on these datasets. Section \ref{sec3} outlines the framework of the proposed method and provides a detailed explanation of its implementation. Section \ref{sec4} describes the model evaluation metrics, compares the performance of our model with other classical models, and makes some discussions. Section \ref{sec5} presents the results of the model application and provides a preliminary analysis of the newly identified EB samples. Section \ref{sec6} summarizes our contributions and discusses the limitations of this study.

\section{Data} \label{sec2}

\subsection{Data description} \label{sec2.1}
TESS conducted high-precision photometry by scanning nearly the entire sky, dividing it into 26 sectors during its two-year Prime Mission. Each sector was observed for approximately 27 days. During this mission phase, TESS collected photometric data at various cadences, including 2-minute LCs for pre-selected targets and 30-minute Full-Frame Images (FFIs). The Extended Mission, which followed the Prime Mission, introduced new data products, such as 10-minute FFIs, further enriching the observational dataset. Target Pixel (TP) files serve as the basis for generating LCs, while Cotrending Basis Vectors (CBVs) are used to correct systematic trends in the data. The most straightforward source of information about a binary system is its photometric LC \citep{vcokina2021automatic}. By analyzing LC, the periodic changes of EBs can be measured and detected, such as eclipse fraction, eclipse depth, eclipse duration, and other parameters, to study the orbital motion and characteristics of stars \citep[e.g.,][]{hurley2002evolution}. Therefore, we chose LCs for analysis in our study and downloaded them from MAST\footnote{\url{https://archive.stsci.edu/missions-and-data/tess}}. Each LC file contains several time series, and PDSAP\_FLUX was selected as the input for our model.

The EBs used for training and testing were selected from the General Catalogue of Variable Stars (GCVS) 5.1 catalog \citep{samus2017general} and the samples published by \cite{prvsa2022tess} from Sectors 1--26 of TESS. We obtained a total of 4,225 EBs. Each star may be observed in multiple sectors, so we have a total of 19,171 LCs for training purposes.

Since our work is a binary classification problem, involving distinguishing data into two predefined categories (e.g., positive or negative classes), constructing negative samples is as important as identifying positive samples. It would be insufficient to consider all negative samples as derived solely from the GCVS 5.1 catalog, as this catalog does not fully represent the variety of non-EBs observed by TESS. The primary goal of our study is to identify EBs from the large set of LCs provided by the TESS mission, which not only observes variable stars but also other astronomical phenomena such as planetary transits, oscillations, flares, stellar rotation, and stellar pulsations. If we restrict the training data to only variable stars, the model's ability to generalize to TESS data would be negatively affected. Therefore, we also include randomly selected LCs from each sector of TESS as negative samples, excluding those known to correspond to EBs. Although some unknown EBs may exist among these randomly selected LCs, their influence is expected to be minimal, as EBs represent only about 2\% to 3\% of TESS’s observed targets. This small proportion of EB samples is unlikely to significantly impact the training outcomes. In total, we collected 18,397 negative samples, including 1,376 from variable stars in the GCVS 5.1 catalog (e.g., BY Draconis variables, $\delta$ Scuti stars, Gamma Doradus variables, M-type stars, RR Lyrae variables of subtype ab, RR Lyrae variables of subtype c, and  Ellipsoidal variables) and 13,760 randomly selected from TESS Sectors 1--74.

\subsection{Data preprocessing} \label{sec2.2}
\begin{figure}[htbp]
\plotone{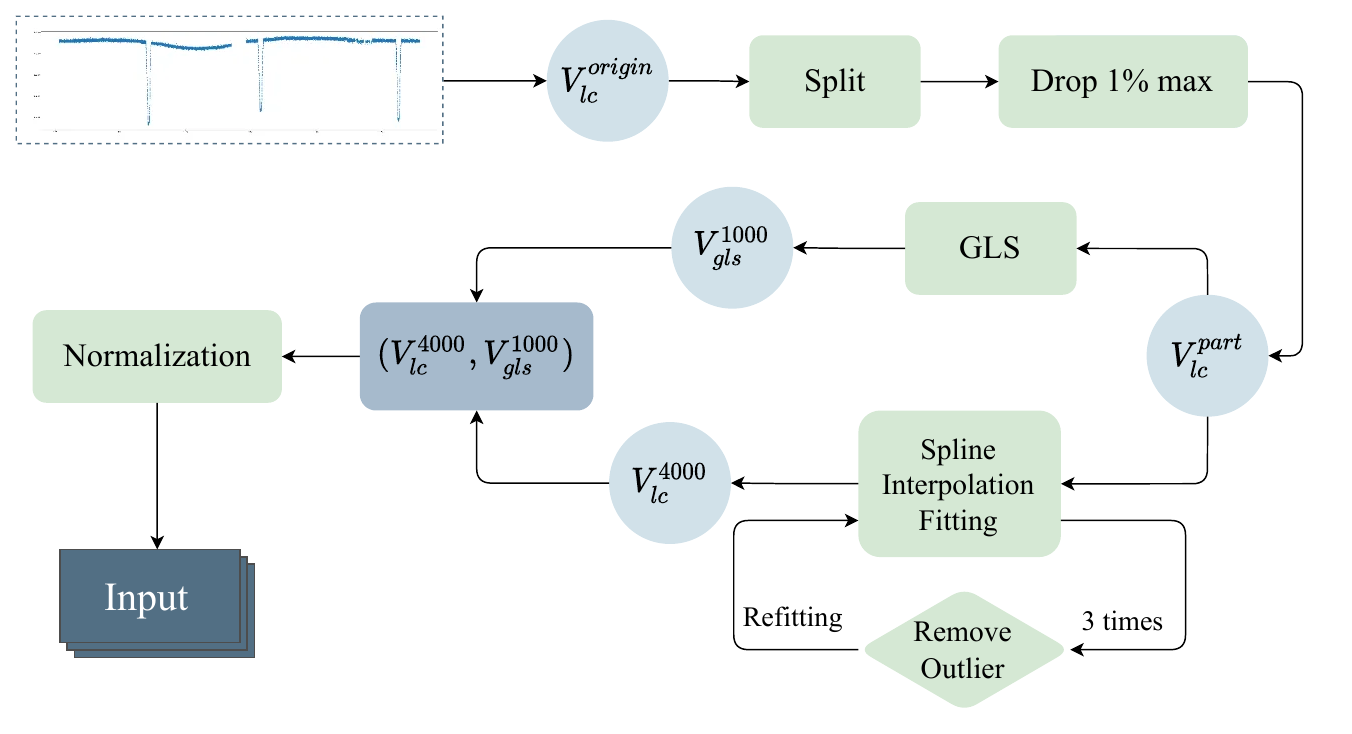}
\caption{Preprocessing flow diagram.
\label{fig1}}
\end{figure}

In recent years, several LC preprocessing methods have focused on calculating the period, such as those by \cite{shallue2018identifying}, \cite{ansdell2018scientific}, and \cite{osborn2020rapid}. These methods typically involve folding each LC within the TCE period to generate 1D vectors as inputs to the neural network. Similarly, \cite{yu2019identifying} used the Box Least Squares (BLS) periodogram to determine the period, followed by phase folding to center the transit. Although these approaches are commonly used in astronomy, they often require extra effort to calculate the period, substantially increasing the workload given the large size of our dataset. Therefore, we designed a simpler preprocessing method using spline interpolation fitting to transform the LC into a vector ($V_{lc}^{4000}$ in Figure \ref{fig1}) that is suitable for input into our neural network model. The overall preprocessing workflow is shown in Figure \ref{fig1}.

\subsubsection{Preprocessing of LC} \label{sec2.2.1}
First, we matched each sample with its corresponding LC, and enhanced our dataset by slicing the LCs. We observed that, due to observational conditions and environmental interference, almost every LC contains blank or missing segments, resulting in discontinuities during certain time periods. To address this, we split each LC at these blank spaces, allowing one LC to be divided into two or three segments. These blank segments are visible in Figure \ref{fig3}. Each segment was treated as an individual sample, increasing the number of positive samples to 36,651 and negative samples to 36,810. The data distribution is summarized in Table \ref{tab1}.

\begin{figure}[htbp]
\plotone{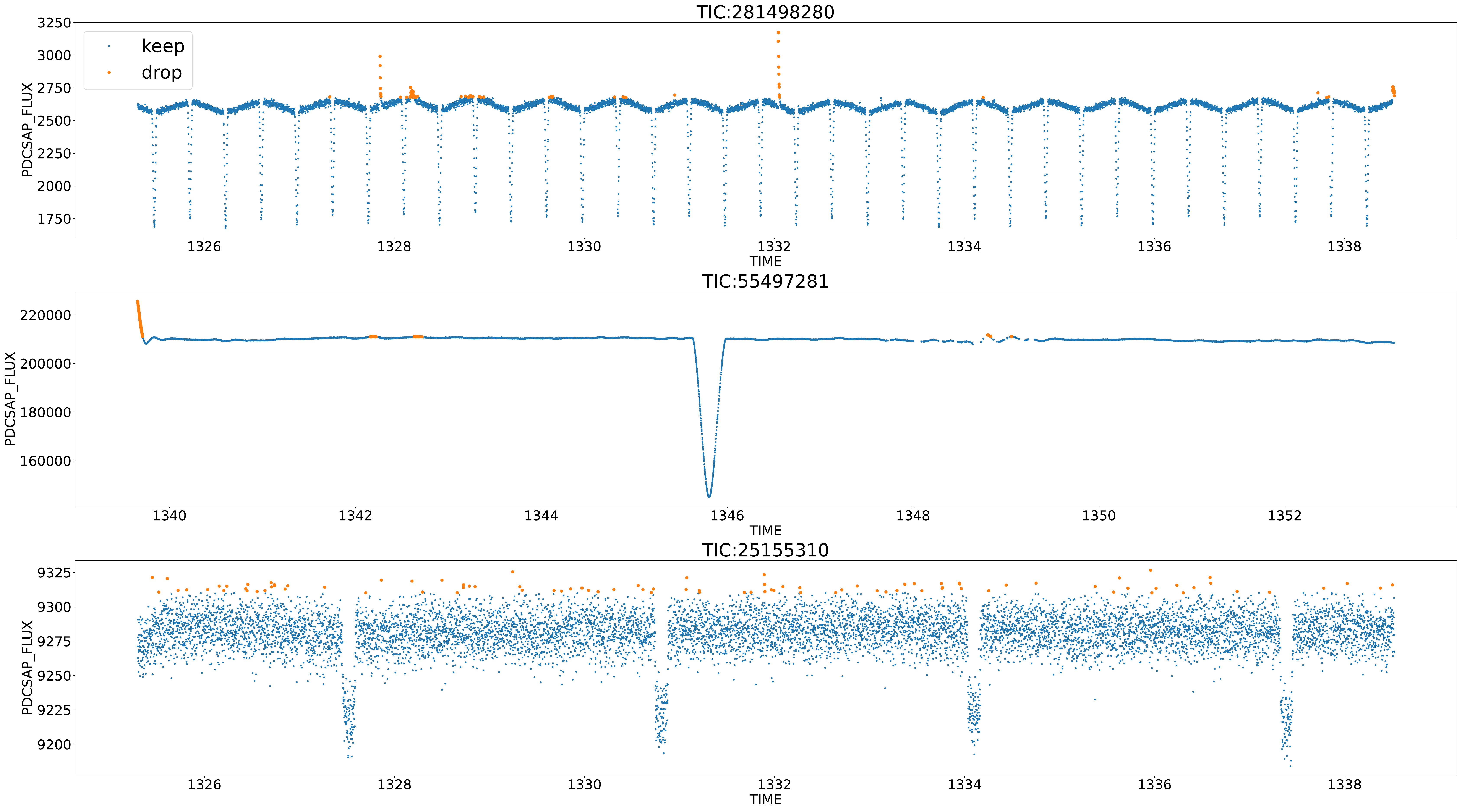}
\caption{Examples of dropping 1\% max flux. Yellow areas represent dropped data points, and blue areas represent retained data points.
\label{fig2}}
\end{figure}

After that, the LCs were processed into a format that the deep learning model can understand. LCs are represented as graphs where the x-axis denotes time and the y-axis denotes flux, illustrating the brightness of an object over a specific period. For each LC, we first removed the top 1\% of flux values to reduce the impact of flares, as shown in Figure \ref{fig2}. Flares are energetic events caused by the reconnection of magnetic field lines \citep{benz2010physical}, resulting in a sudden increase in flux. We only applied top-tail truncation (top 1\% of flux values) rather than bottom-tail truncation, as the flux decreases sharply during eclipses, a key feature of EBs. The output of this step was denoted as $V_{lc}^{part}$, as shown in Figure \ref{fig1}.

\begin{table}[ht]
\renewcommand{\arraystretch}{1.5}
\setlength{\tabcolsep}{10pt}
\centering
\caption{Data Distribution}
\label{tab1}
\begin{tabular}{cccc}
\hline\hline
    \textbf{Sample Type}  & \textbf{Raw LC amount} &\textbf{Split LC amount} &\textbf{GLS amount} \\ \hline
    Positive(EBs) & 19171 & 36651 & 36651  \\ \hline
    Negative(Non-EBs) & 18397 & 36810 & 36810         \\ \hline
\end{tabular}
\end{table}

Next, we adopted spline interpolation to standardize the LCs of varying lengths to a uniform one. We observed that the lengths of $V_{lc}^{part}$ generated from different LCs ranged from 2000 to 9000 data points. To balance model complexity and feature preservation, we interpolated the length to 4000. If the input length is excessively long, it increases the complexity of the processing work; whereas if it is too short, it may suffer from insufficient feature information, resulting in inaccurate training outcomes.

To address offsets and inconsistencies between different sectors of the same star, we employed an iterative spline fitting and normalization approach to preprocess the LC data. First, we applied a cubic spline fitting to the raw LC data. The spline was fitted to capture the overall trend of the LC while smoothing out noise and irregularities. Following the initial spline fit, we identified outliers as data points that deviated by more than $3\sigma$ from the fitted curve. Subsequently, we removed these outliers and refitted the spline curve using the remaining data \citep{vanderburg2014technique}. This iterative process—outlier identification, removal, and spline refitting—was repeated three times. Once the iterative spline fitting was completed, we normalized the flux values to adjust the LC to a consistent scale across all sectors. This normalization was achieved using formula (\ref{eq1}), where the flux values were rescaled to account for baseline variations and amplitude differences. The resulting preprocessed LC data, denoted as $V_{lc}^{4000}$, retained the essential characteristics of the LC while reducing sector-specific shifts and inconsistencies. This entire preprocessing workflow is visually represented in Figure \ref{fig1}. Formula (\ref{eq1}) is defined as follows:
\begin{equation}
y_{i} = \frac {x_{i} - \min_{1\leq j \leq n}{\{x_{j}\}}} {\max_{1\leq j \leq n}{\{x_{j}\}} - \min_{1\leq j \leq n}{\{x_{j}\}}}
\label{eq1}
\end{equation}
\\

\subsubsection{Preprocessing of GLS} \label{sec2.2.2}

\begin{figure}[htbp]
\plotone{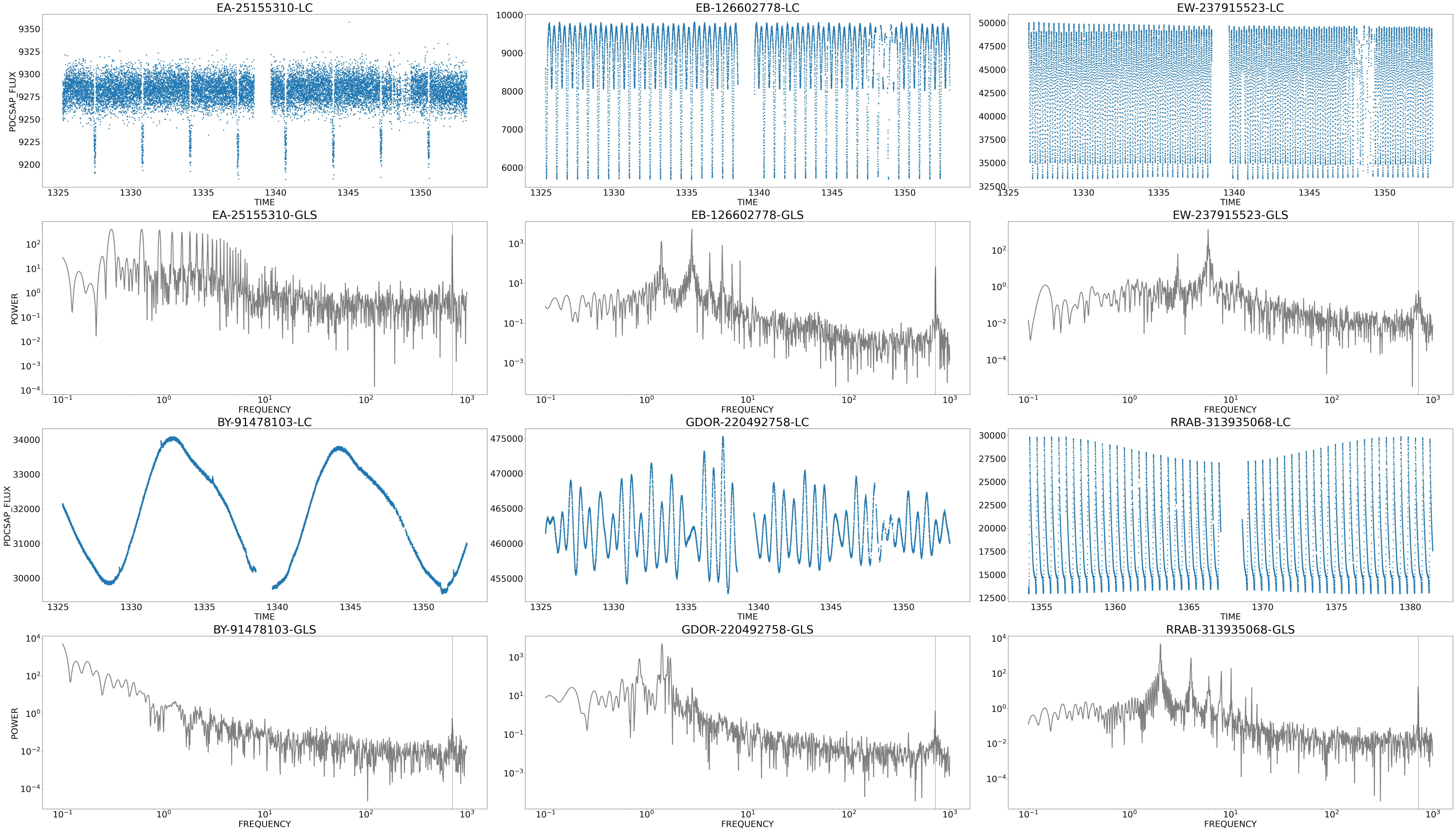}
\caption{Examples of TESS 2-minute cadence light curves (plotted in blue) of six different types of variables (from upper left to bottom right: EA, EB, and EW types of eclipsing binaries, BY Draconis, $\gamma$ Doradus, and RR Lyrae-ab variables) and their corresponding GLS periodograms (plotted in gray). Data in the first two rows are included in the positive sample, while the rest are included in the negative sample.}
\label{fig3}
\end{figure}

In order to enrich our data features, we employed GLS \citep{zechmeister2009generalised}, a technique designed to detect periodic signals in non-uniformly spaced discrete data points, to extract the power spectrum of periodic signals from LCs. This characteristic is significant for the analysis of EBs, as it aids in uncovering physical properties such as periodic variations and orbital parameters. We applied the GLS method to $V_{lc}^{part}$ as illustrated in Figure \ref{fig1}. The calculation of GLS can be found in \cite{zechmeister2009generalised}. We set the length of GLS data to 1000 to facilitate the subsequent fusion with LC data. The GLS data is displayed in Figure \ref{fig3}, with the x-axis representing frequency and the y-axis representing the corresponding power spectral density. Power spectral density reflects the strength or importance of signal energy. Subsequently, we applied max-min normalization (Formula \ref{eq1}) on the y-axis data. The resulting $V_{gls}^{1000}$ was then used as one of the inputs of our model.

Following the preprocessing steps outlined above, we obtain data pairs $(V_{lc}^{4000}, V_{gls}^{1000})$ which can be input into our deep learning models. There exists a one-to-one correspondence between the LC data and the GLS data, indicating that the number of GLS data points is equal to the number of LC data points. The distribution of the GLS data is presented in Table \ref{tab1}. We can observe the plots of some samples for both LC and GLS data in Figure \ref{fig3}.

\section{Method} \label{sec3}

\subsection{Framework of method} \label{sec3.1}
\begin{figure}[ht!]
\plotone{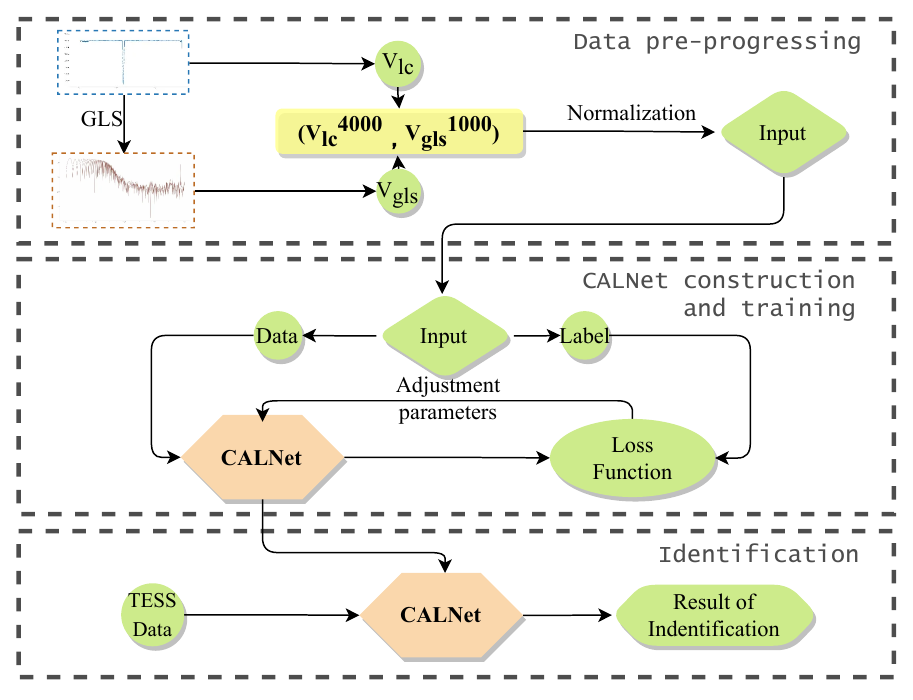}
\caption{Framework of our method. It comprises the data processing, model construction and training, and identification of new EBs.}
\label{fig4}
\end{figure}

The overall framework of the method for identifying EBs using TESS data is introduced below, as shown in Figure \ref{fig4}. The first part of the method is the preprocessing of the data, which we have described in Section \ref{sec2.2}. Then the second part is the core of our method – construction and training of CALNet, which uses the processed data pairs $(V_{lc}^{4000},V_{gls}^{1000})$ as input and will be described in detail in Section \ref{sec3.2}. The third part is classification. We put all the TESS data we have (from Sectors\,1--74) into the trained CNN\_Attention\_LSTM Net (CALNet) for classification and identifying new EBs. The classification result will be evaluated in Section \ref{sec4} and discussed in Section \ref{sec5}.

\subsection{Model construction} \label{sec3.2}
\subsubsection{Baselines construction} \label{sec3.2.1}
In order to verify the feasibility and superiority of our CALNet, we first built a basic CNN model using only the LC data. We also introduced several state-of-the-art classification models for LCs, including AlexNet \citep{krizhevsky2012imagenet}, AstroNet \citep{shallue2018identifying}, RLNet \citep{yan2023feature} and biLSTM+CNN \citep{vcokina2021automatic}. These models serve as baselines for comparison and analysis with our CALNet. The details of the basic CNN model are provided below, while the descriptions of the other models are presented in Section \ref{sec4.2.3}.

\begin{figure}[ht!]
\plotone{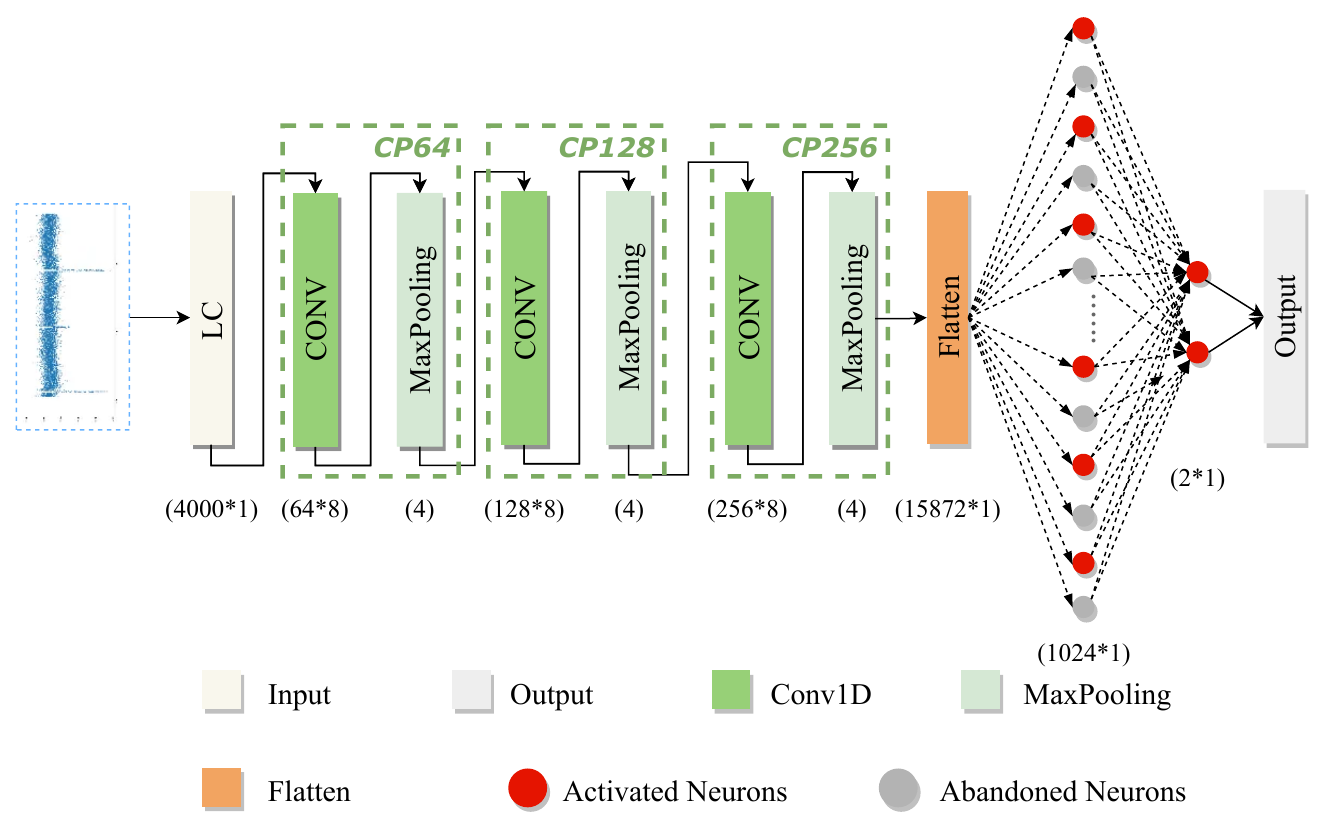}
\caption{Basic CNN model. Different modules are distinguished using different colors.}
\label{fig5}
\end{figure}

Our basic CNN model is designed with a single input, LCs. Figure \ref{fig5} illustrates the architecture of our CNN model, which consists of the following components:

\begin{enumerate}[(1)]
    \item Input Layer. The input to this layer is $V_{lc}^{4000}$, derived from the preprocessing step in Section \ref{sec2.2.1}.
    \item Three Convolutional-Pooling (CP) Modules. Each CP module consists of a one-dimensional convolution layer followed by a pooling layer. The numbers of convolutional filters in the three CP modules are 64, 128, and 256, respectively, denoted as CP64, CP128, and CP256. Features are first extracted by convolutional layers (filters(64, 128, or 256)$\times$kernel size(8)), and ReLU activation functions (Formula\,\ref{eq2}) are added to each convolutional layer to enable neurons to perform linear transformations and produce nonlinear outputs. 
    \begin{equation}
    \text{ReLU} = \max(0, x)
    \label{eq2}
    \end{equation}
    Then outputs are passed through a Maxpooling layer (pool size = 4), which selects the maximum value from each local region, enhancing feature extraction. Through the three CP modules, we extract increasingly complex features from the LCs.
    \item Flatten Layer. The data is then flattened into a $15872\times1$ vector, which serves as input to the subsequent fully connected layer.
    \item Two Fully Connected Layers. The first fully connected layer contains 1024 neurons, with a sigmoid (Formula \ref{eq3}) activation function to introduce nonlinearity and increase model expressiveness. The second fully connected layer uses the Softmax activation function to perform the final classification. The Softmax (Formula\,\ref{eq4})  function computes the prediction probabilities for each category based on the features obtained from the previous layer, helping determine the class of each LC.
    \begin{equation}
    S(x) = \frac{1}{1+e^{-x}}
    \label{eq3}
    \end{equation}
    \begin{equation}
    Softmax(x) = \frac{e^{x_{i}}}{\sum_{i}e^{x_{i}}}
    \label{eq4}
    \end{equation}
    \item Output Layer. The output layer produces a two-dimensional vector, where each column represents the predicted probability for each category (i.e., positive or negative). 
\end{enumerate}

\subsubsection{CALNet construction} \label{sec3.2.2}
Focusing on our data pairs (LC, GLS), we construct a hybrid deep learning model called CNN\_Attention\_LSTM Net (CALNet), which incorporates both an Attention Module and Long Short-Term Memory (LSTM) to enhance the robustness of the CNN model. For the attention mechanism, we utilize the Convolutional Block Attention Module (CBAM), a neural network component widely used in computer vision tasks. CBAM adaptively extracts features from the input data within the convolutional network and assigns appropriate weights to these features, improving the network’s performance in tasks such as classification, detection, and segmentation \citep{woo2018cbam}. Additionally, we incorporate LSTM to address the challenges of vanishing and exploding gradients. By introducing gate functions into the cell structure, LSTM effectively controls the flow of information, enabling the model to capture long-term dependencies in sequential data \citep{sherstinsky2020fundamentals}.

The CALNet structure is described in detail as follows and shown in Figure \ref{fig6}:

\begin{figure}[htbp]
    \centering
    \includegraphics[width=0.9\linewidth]{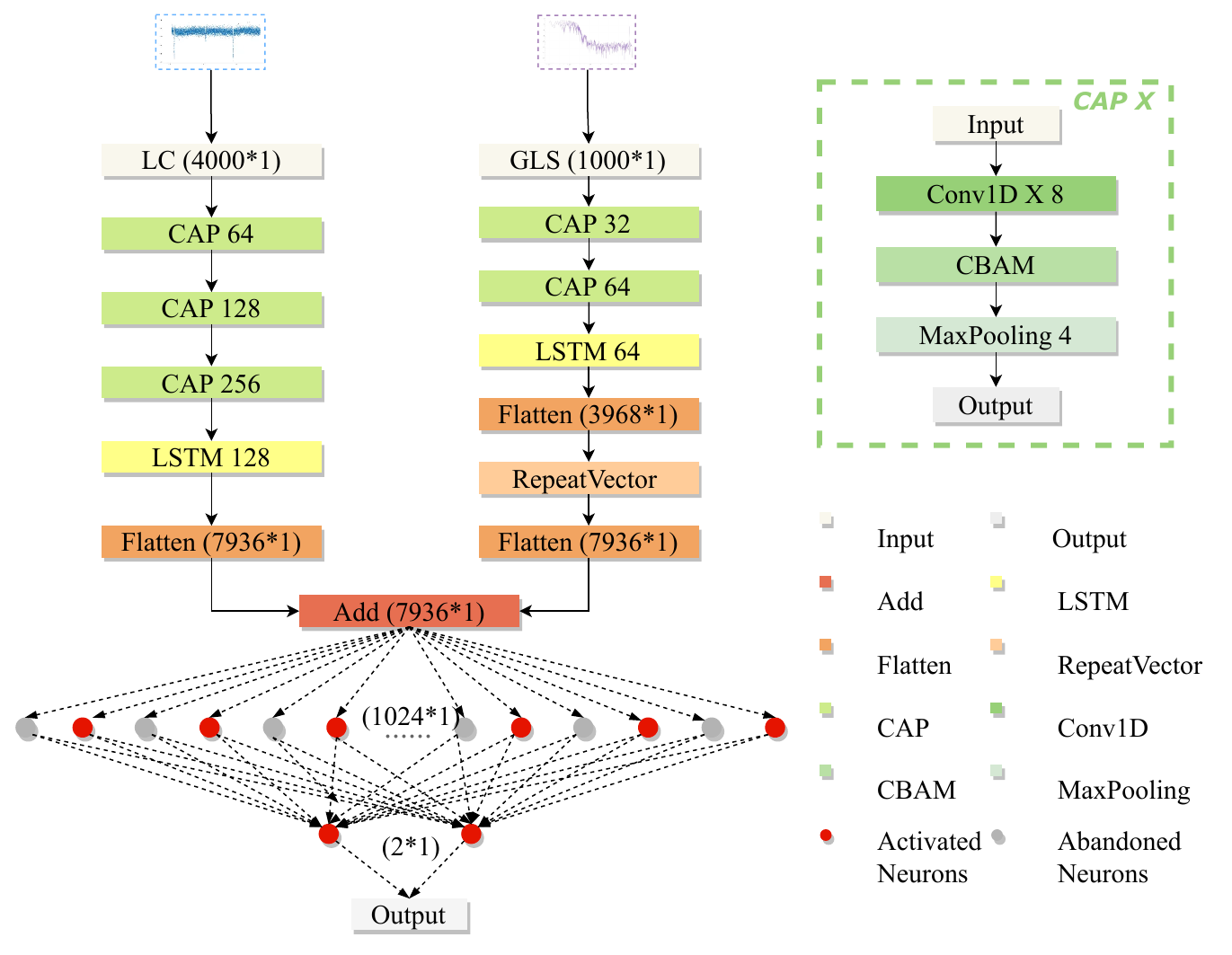}
    \caption{Architecture of CALNet. The structure of CAP modules is shown in the right panel.}
    \label{fig6}
\end{figure}

\begin{enumerate}[(1)]
    \item LC Branch
    \begin{enumerate}[(i)]
        \item Input Layer. It takes the processed LC data as input. 
        \item Three CAP Modules. Each module consists of a one-dimensional convolution layer, an Attention block, and a max-pooling layer. The filter sizes in the convolution layers are 64, 128, and 256, respectively, resulting in the modules being denoted as CAP64, CAP128, and CAP256. The convolution and max-pooling layers have the same parameters as those in the basic CNN model.
        \item LSTM Layer. This layer consists of 128 neuron units, capturing sequential dependencies in the data.
        \item  Flatten Layer. The data is flattened into a $7936\times1$ vector to prepare for the subsequent fully connected layers.
    \end{enumerate}
    
    \item GLS Branch
    \begin{enumerate}[(i)]
        \item Input Layer. 
        \item Two CAP Modules. These modules are CAP32 and CAP64, with similar structures to the LC branch but with different filter sizes.
        \item LSTM Layer. This layer contains 64 units to capture sequential dependencies in the GLS data.
        \item Flatten Layer. The output is flattened into a $3968\times1$ vector.
        \item RepeatVector Layer. The $3968\times1$ vector is duplicated to create a $3968\times2$ vector, enabling subsequent fusion of the multi-type data.
        \item Flatten Layer. The $3968\times2$ vector is flattened into a one-dimensional $7936\times2$ vector for fusion with the LC branch.
    \end{enumerate}
    
    \item Add Layer. The outputs from both branches (each $7936\times2$) are fused by element-wise addition, resulting in a $7936\times2$ vector.
    \item Fully Connected Layers. Two fully connected layers follow, with the same parameter settings as those in the basic CNN model.
    \item Output Layer. The final output layer provides the binary classification prediction.
\end{enumerate}

\begin{figure}[ht!]
    \plotone{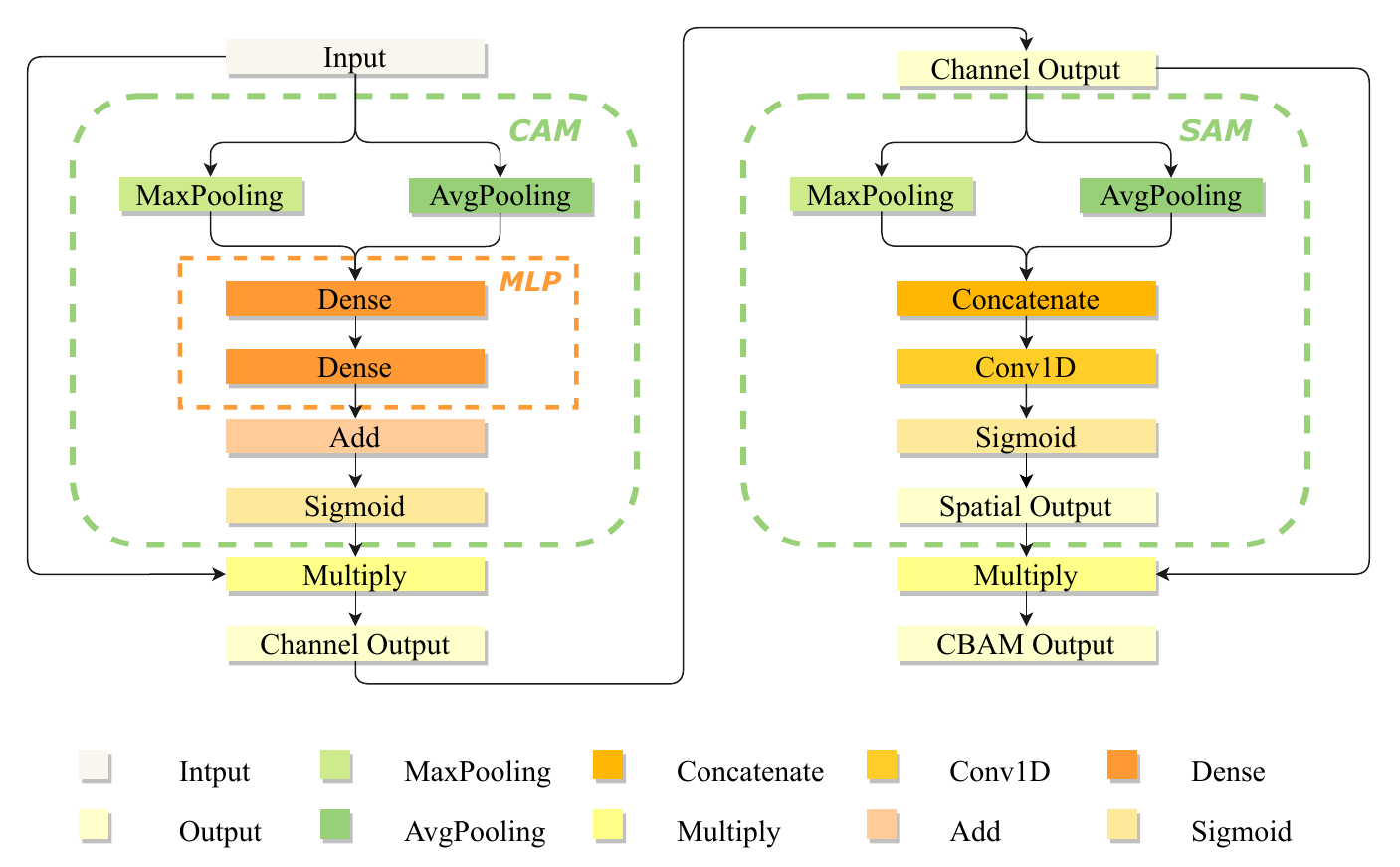}
    \caption{Architecture of CBAM. On the left is the CAM, and on the right is the SAM.}
    \label{fig7}
\end{figure}

The detailed structure of the CBAM is shown in Figure \ref{fig7}. It consists of two submodules: the Channel Attention Module (CAM) and the Spatial Attention Module (SAM). CAM performs nonlinear mapping and compression by pooling global information into a vector, followed by a series of fully connected layers. This allows the module to learn the importance weights for each channel in the feature map, enabling the model to focus more on features that are critical for the task at hand. SAM captures contextual information at various scales by performing spatial pooling operations of different proportions on the feature maps. These pooled features are then fused using convolution operations and a sigmoid activation function, which results in an efficient spatial attention map. CBAM combines CAM and SAM in series, allowing it to consider both channel-wise and spatial-wise characteristics simultaneously. This attention mechanism effectively suppresses noise and irrelevant features while preserving important ones, thereby enhancing the expressive power and performance of the network.

\subsection{Model training} \label{sec3.3}
We divided all the samples into training and testing sets in an 80:20 ratio, randomly selecting 80\% of the data for the training set and the remaining 20\% for the testing set. The data distribution is shown in Table \ref{tab2}. To mitigate the risk of overfitting, we incorporated a dropout function in the fully connected layers. This randomly removes 50\% of the neurons with a probability of 0.5 during training, helping to prevent excessive reliance between units and reducing co-adaptation \citep{srivastava2014dropout}.

For the loss function, we use cross-entropy loss (formula (\ref{eq5})) suitable for binary classification. The optimization process is handled by the Adam optimizer \citep{kingma2014adam}, which minimizes the cross-entropy error function on the training set. The model is trained with a batch size of 128, a learning rate of $\alpha = 10^{-3}$, with a total of 30 training epochs.
\begin{equation}
    Loss = -\frac{1}{N}\sum_{i=1}^{N} y_{i}\cdot \log(p(y_i))+(1-y_i)\cdot\log(1-p(y_i)) 
\label{eq5}
\end{equation}

\begin{table}[!h]
\renewcommand{\arraystretch}{1.5}
\setlength{\tabcolsep}{10pt}
\centering
\caption{Data Distribution of Training and Testing Set}
\label{tab2}
\begin{tabular}{ccc}
\hline\hline
    \textbf{Data Amount}  & \textbf{Positive class} &\textbf{Negative class} \\ \hline
    Training set & 29327 & 29441   \\ \hline
    Testing set & 7324 & 7369         \\ \hline
\end{tabular}
\end{table}

\section{Model evaluation and performance}  \label{sec4}

\subsection{Evaluation metrics}  \label{sec4.1}
For our classification, the results are summarized in a confusion matrix, as shown in Table \ref{tab3}. The confusion matrix is a fundamental and intuitive method for evaluating the accuracy of a classification model \citep{sokolova2009systematic}. In this matrix, TP (True Positive) indicates the number of positive samples correctly predicted by the model, FN (False Negative) represents the number of positive samples incorrectly predicted as negative, FP (False Positive) indicates the number of negative samples incorrectly predicted as positive, and TN (True Negative) shows the number of negative samples correctly predicted by the model.

\begin{table}[!h]
\renewcommand{\arraystretch}{1.5}
\setlength{\tabcolsep}{6pt}
\centering
\caption{Confusion Matrix}
\label{tab3}
\begin{tabular}{ccc}
\hline\hline
             & Predicted True & Predicted False \\ \hline
Actual True  & TP             & FN              \\ \hline
Actual False & FP             & TN              \\ \hline
\end{tabular}
\end{table}

Referring to the classical evaluation metrics for model prediction \citep{sokolova2009systematic}, we employ Accuracy, Precision, Recall, and F1-score to evaluate our models. Using the confusion matrix, the formulas for calculating these metrics are as follows:
\begin{equation}\label{eq6}
    Accuracy=\frac{TP+TN}{TP+TN+FP+FN}
\end{equation}
\begin{equation}\label{eq7}
    Precision=\frac{TP}{TP+FP}   
\end{equation}
\begin{equation}\label{eq8}
    Recall=\frac{TP}{TP+FN}    
\end{equation}
\begin{equation}\label{eq9}
    F1-score= \frac{2\times Precision\times Recall}{Precision+Recall}
\end{equation}

\subsection{Comparison and analysis}  \label{sec4.2}
\subsubsection{Comparison of different feature fusion methods}  \label{sec4.2.1}
In this section, we evaluate the impact of three feature fusion methods on CALNet. Feature fusion refers to the process of combining features extracted from different data sources or modalities to enhance the model's ability to capture complementary information. In our study, we use LC and GLS data as inputs to two separate network branches, enabling targeted feature extraction for each data type. Common feature fusion methods include addition (add), which combines features by element-wise summation; concatenation (concatenate), which stacks features along a specific dimension; and multiplication (multiply), which fuses features through element-wise interaction. To analyze the performance of these methods, we designed three models consistent with the structure described in Section \ref{sec3.2.2}, modifying only the feature fusion strategy. The results of these models are summarized in Table \ref{tab4}.

\begin{table}[!h]
\renewcommand{\arraystretch}{1.5}
\setlength{\tabcolsep}{10pt}
\centering
\caption{Confusion Matrix}
\label{tab4}
\begin{tabular}{lcccc}
\hline \hline
    \textbf{Fusion methods} &\textbf{Accuracy} &\textbf{Precision} &\textbf{Recall} & \textbf{F1-score} \\ \hline
    Add            & \textbf{0.965} & 0.957 & \textbf{0.972} & \textbf{0.965}          \\ \hline
    Concatenate     & 0.956 & 0.950 & 0.961 & 0.956             \\ \hline
    Multiply        & 0.958 & \textbf{0.961} & 0.954 & 0.958                \\ \hline
\end{tabular}
\tablecomments{Bold values indicate the highest performance for each metric across the fusion method.}
\end{table}

According to Table \ref{tab4}, there is no significant difference in the model's performance across the three data fusion methods. However, the model's performance is relatively superior when using the Add fusion method. Therefore, we chose the addition method for data fusion in our model for subsequent work.

\subsubsection{Comparison of the different factors in our models}  \label{sec4.2.2}
\begin{figure}[ht!]
\includegraphics[width=\linewidth]{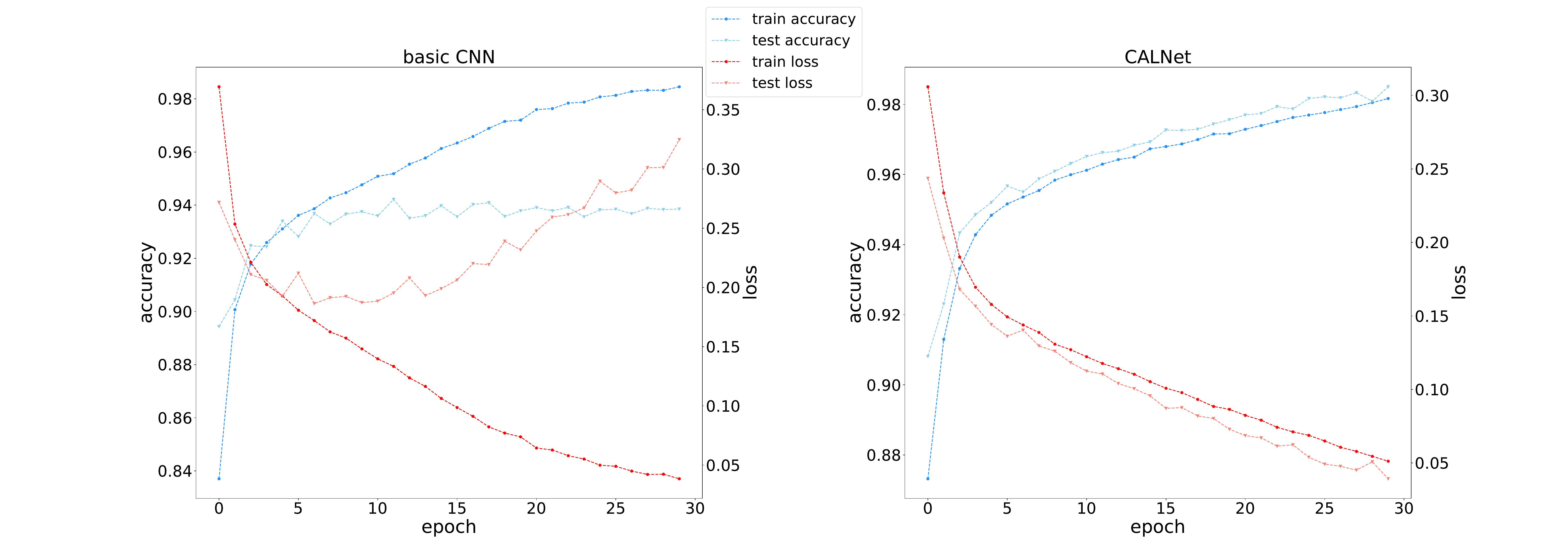}
\caption{Accuracy-loss curves of basic CNN and CALNet on training set and test set.
\label{fig8}}
\end{figure}

In this section, we compare the performance of the basic CNN model in Section \ref{sec3.2.1} and CALNet model in Section \ref{sec3.2.2}, and provide a detailed analysis of the factors contributing to the differences in their performance. First, we plotted the accuracy and loss function changes for both the basic CNN and CALNet on the training and the test sets, as shown in Figure \ref{fig8}. 

\begin{figure}[ht!]
    \centering
    \begin{minipage}{0.49\linewidth}
        \centering
        \includegraphics[width=\linewidth]{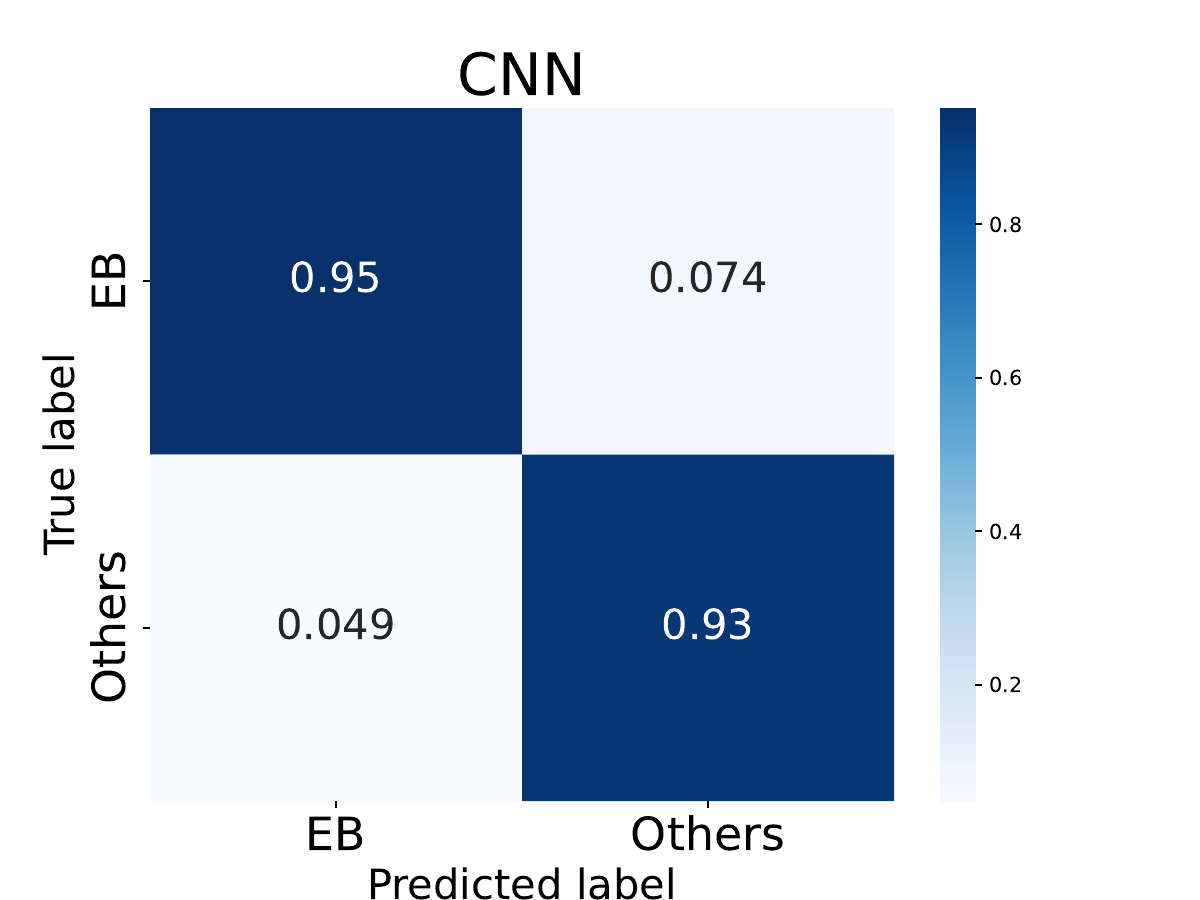}
    \end{minipage}
    \begin{minipage}{0.49\linewidth}
        \centering
        \includegraphics[width=\linewidth]{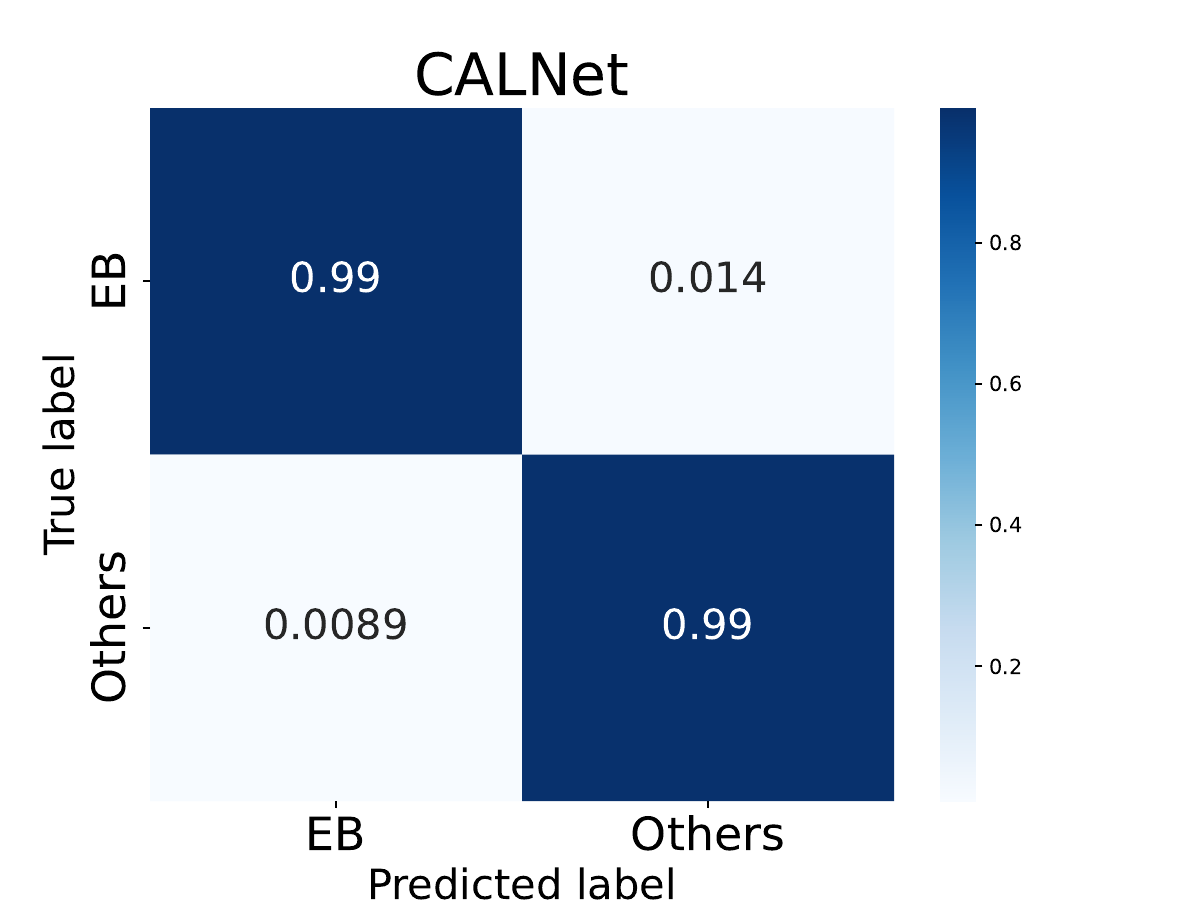}
    \end{minipage}
    \caption{Confusion matrix of basic CNN  and CALNet.
\label{fig9}}
\end{figure}

As can be seen from Figure \ref{fig8}, CALNet demonstrates better convergence than the basic CNN. On the training set, the basic CNN achieves an accuracy of around 0.98, with the loss dropping below 0.05. However, on the test set, the performance of the basic CNN reveals signs of overfitting, as it struggles to generalize well to the new test data. In contrast, CALNet exhibits a consistent trend in both accuracy and loss across the training and test sets, indicating better generalization. The confusion matrix corresponding to the test set results for both the basic CNN and CALNet is shown in Figure \ref{fig9}, clearly demonstrating the superior performance of CALNet.

To make our work more comprehensive and rigorous, we also conducted several ablation experiments. Compared to the base CNN, our CALNet incorporates an additional data type (GLS), the CBAM attention module, and the introduction of LSTM. We separately evaluated the impact of each of these three factors on model performance. The results are shown in Table \ref{tab5}, from which we can derive the following conclusions:

\begin{table}[ht]
\setlength{\tabcolsep}{10pt}
\centering
\caption{Influences of Data, CBAM and LSTM on the Model Effect Respectively}
\label{tab5}
\resizebox{\textwidth}{!}{
\begin{tabular}{ccccccc}
\hline\hline
    \multicolumn{3}{c}{\textbf{Factors}}    &\multicolumn{4}{|c}{\textbf{Metrics}}   \\ \hline
    Data &CBAM &LSTM &Accuracy &Precision &Recall &F1-score   \\ \hline
    \textbf{LC only} 
    & -             & -             & \multicolumn{1}{|c}{0.944}    & 0.940 & 0.938 & 0.939  \\ 
    & \checkmark    & -             & \multicolumn{1}{|c}{0.946}    & 0.944 & 0.936 & 0.940 \\  
    & -             & \checkmark    & \multicolumn{1}{|c}{0.964}    & 0.966 & 0.950 & 0.958 \\ 
    & \checkmark    & \checkmark    & \multicolumn{1}{|c}{\textbf{0.967}}    & \textbf{0.971} & \textbf{0.958} & \textbf{0.964} \\ \hline
    GLS only
    & -             & -             &\multicolumn{1}{|c} {0.936}    & 0.935 & 0.919 & 0.927 \\
    & \checkmark    & -             &\multicolumn{1}{|c} {0.940}    & 0.936 & 0.933 & 0.934 \\
    & -             & \checkmark    &\multicolumn{1}{|c} {0.958}    & \textbf{0.963} & 0.955 & \textbf{0.959} \\
    & \checkmark    & \checkmark    &\multicolumn{1}{|c} {\textbf{0.962}}    & 0.950 & \textbf{0.960} & 0.955 \\ \hline
    LC + GLS
        & -             & -             & \multicolumn{1}{|c}{0.968}    & 0.962 & 0.972 & 0.967 \\
        & \checkmark    & -             & \multicolumn{1}{|c}{0.972}    & 0.972 & 0.980 & 0.976 \\
        & -             & \checkmark    & \multicolumn{1}{|c}{0.980}    & 0.978 & 0.984 & 0.981 \\
        & \checkmark    & \checkmark    & \multicolumn{1}{|c}{\textbf{0.989}}    & \textbf{0.986} & \textbf{0.991} & \textbf{0.989} \\ \hline 
\end{tabular}
}
\tablecomments{Bold values highlight the best performance for each metric column. The first row shows baseline CNN results, while the last row presents our proposed CALNet's performance.}
\end{table}

(1) Data Fusion: The model that combines multi-type data (LC and GLS) outperforms the model using a single data type. Specifically, the model using only LC data achieves a higher F1-score than the one using GLS alone. This can be attributed to the fact that LC data captures important features such as luminosity variations and periodicity, while GLS mainly reflects spectral characteristics. Additionally, LC data has a higher dimensionality (4,000) compared to GLS (1,000), providing more information overall. 

(2) Effect of CBAM: Introducing the CBAM improves model performance, particularly when dealing with multi-type data. As shown in Table~\ref{tab5}, the inclusion of CBAM positively impacts both the single-LC and single-GLS models. However, the improvements in all four evaluation metrics are more pronounced when CBAM is applied to the multi-type data model, indicating that CBAM enhances the model's ability to focus on crucial features.

(3) Impact of LSTM: The incorporation of LSTM further enhances model performance, as evidenced by improvements in all four evaluation metrics. The gains from LSTM are more significant than those from CBAM. This suggests that, since our data is time-series-based, LSTM’s ability to capture long-term dependencies makes it particularly effective in processing time-dependent data, highlighting the necessity of LSTM in our model.

\subsubsection{Comparison with other models}  \label{sec4.2.3}
We compare CALNet with several classification models for LCs, including AlexNet \citep{krizhevsky2012imagenet}, AstroNet \citep{shallue2018identifying}, RLNet \citep{yan2023feature}, and biLSTM+CNN \citep{vcokina2021automatic}.

AlexNet \citep{krizhevsky2012imagenet} is a deep convolutional neural network model that was applied on 1.2 million high-resolution images from ImageNet LSVRC-2012 contest to achieve classification, achieving top-1 and top-5 error rates of 37.0\% and 17.0\% on the test dataset.

BiLSTM+CNN \citep{vcokina2021automatic} is a deep learning model to classify EBs. It was trained using 491,425 synthetic LC data generated by ELISa software for training and evaluated on 100 observed binary star LCs, achieving 98\% accuracy on the evaluation set. 

RLNet \citep{yan2023feature} is an LC classifier designed for Kepler data. It classifies approximately 150,000 Kepler LCs into 11 categories with an accuracy of 98.7\%, a minimum recall of 96.8\%, and a minimum precision of 97.0\% under all categories.

AstroNet \citep{shallue2018identifying} is a deep convolutional neural network for identifying whether a signal is a transiting exoplanet or a false positive caused by astrophysical or instrumental phenomena. AstroNet achieved 96\% accuracy and an AUC of 98.8\%. The test data consisted of 1523 TCEs from the Autovetter Planet Candidate Catalog (Q1--Q17 DR24) hosted at the NASA Exoplanet Archive.

\begin{figure}[ht!]
\centerline{\includegraphics[width=0.6\linewidth]{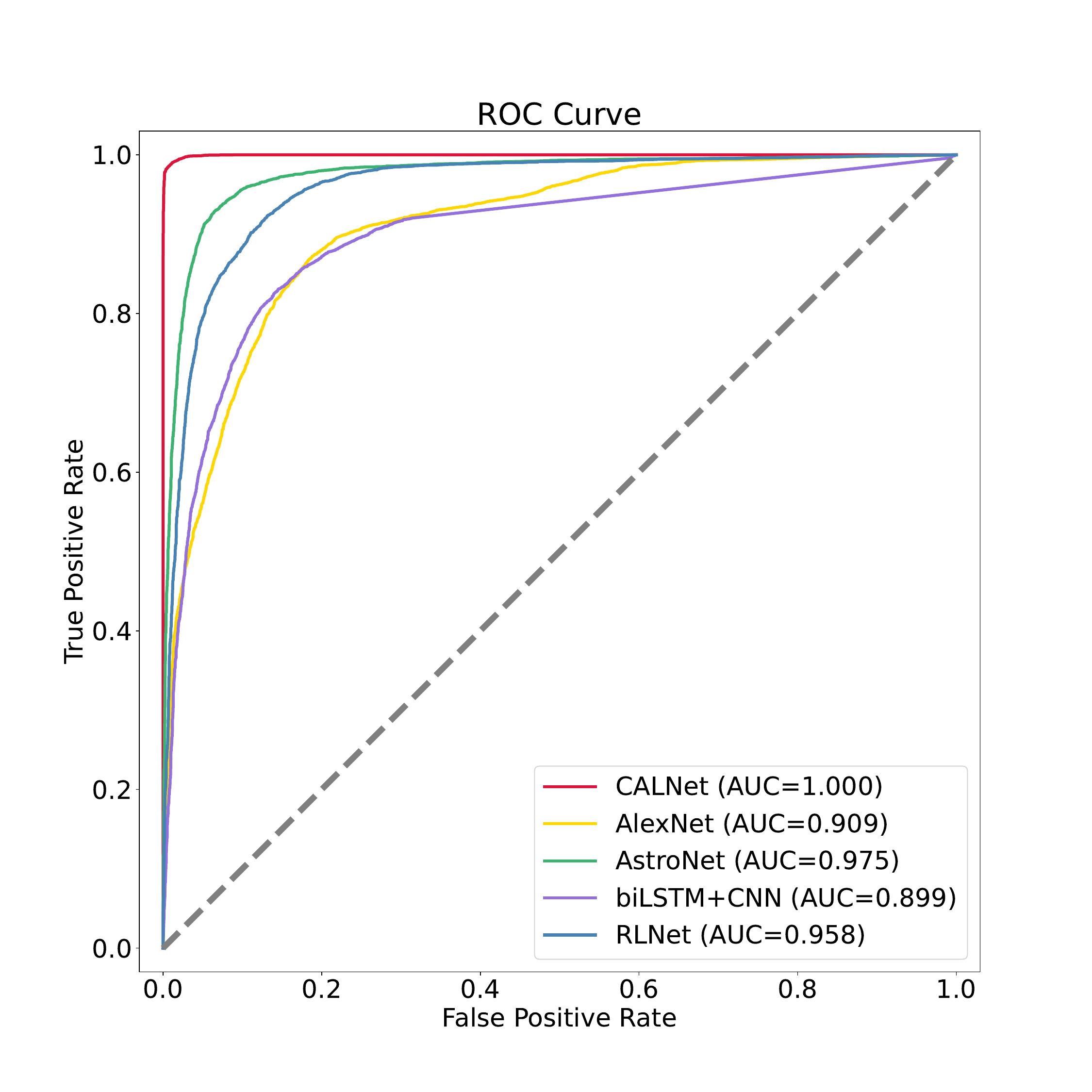}}
\caption{ROC curve. The x-axis represents the False Positive Rate (FPR), the y-axis represents the True Positive Rate (TPR), and the Area Under the Curve (AUC) value is displayed in the lower right corner.}\label{fig10}
\end{figure}

The first three models mentioned above require only a single input. For this analysis, we use LC data as their input since our previous analysis (as shown in Table \ref{tab5}) indicates that LC data consistently outperforms GLS data. The last model, which requires two inputs, utilizes our data pair (LC, GLS) for training and testing, the same as CALNet. To ensure fairness in comparison, we conducted parameter tuning for these four models before evaluating their performance. All models were executed under the same computational environment. The performance metrics for each model are presented in Table \ref{tab6}, and the ROC curve for these models is shown in Figure \ref{fig10}.

\begin{table}[!h]
\renewcommand{\arraystretch}{1.5}
\setlength{\tabcolsep}{10pt}
\centering
\caption{Performance of Each Model}
\label{tab6}
\begin{tabular}{lcccc}
\hline \hline
    \textbf{Model/Metrix} &\textbf{Accuracy} &\textbf{Precision} &\textbf{Recall} & \textbf{F1-score}                  \\ \hline
    AlexNet (2012)      & 0.843 & 0.830 & 0.851 & 0.840     \\ \hline
    biLSTM + CNN (2021) & 0.840 & 0.881 & 0.814 & 0.846     \\ \hline
    RLNet (2023)        & 0.926 & 0.932 & 0.935 & 0.933     \\ \hline
    AstroNet (2019)     & 0.951 & 0.958 & 0.954 & 0.956     \\ \hline
    CALNet (this work)  & \textbf{0.989} & \textbf{0.986} & \textbf{0.991} & \textbf{0.989}            \\ \hline
\end{tabular}
\tablecomments{Bold values indicate the best performance in each metric column.}
\end{table}

The ROC curve in Figure \ref{fig10} evaluates classification model performance by plotting True Positive Rate (TPR) (Sensitivity/Recall) against False Positive Rate (FPR) across thresholds. TPR measures the ability to correctly identify positives. FPR indicates the proportion of negatives incorrectly classified as positives. The AUC (Area Under the Curve) quantifies performance, with values closer to 1 indicating better classification.

From Table \ref{tab6} and Figure \ref{fig10}, it is evident that our model, CALNet, outperforms the other four models. Here are some analyses for this result. AlexNet was designed primarily for image-type data and developed to handle datasets like ImageNet. Since our work involves LCs, which are time-series data, it may not be well-suited for this task. AstroNet was originally created to detect TCEs in Kepler data \citep{shallue2018identifying}; it has also been confirmed to work with LC data from TESS \citep{yu2019identifying}. In their study, AstroNet utilized two inputs—global view and local view, both preprocessed from LCs and subjected to phase-folding, which aligns and centers the transit events. These inputs had lengths of 201 and 61, respectively. In contrast, our two-input data are derived from LC and GLS data, without phase-folding, and have lengths of 4,000 and 1,000, respectively. This difference in preprocessing and data structure may hinder AstroNet's ability to handle our complex, multivariate dataset effectively. For biLSTM + CNN, this model was designed for classifying EBs. However, in the authors' work, it was trained exclusively on simulated LCs with a length of only 100 points. In contrast, our task involves real LCs with significantly longer sequences, making this model less adaptable to our dataset. For RLNet, it is well-suited for Kepler LCs, which span up to 4 years and contain approximately 70,000 data points, but LCs from TESS differ significantly. TESS data typically cover only $\sim 27$ days and have far fewer data points \citep{shallue2018identifying}. This disparity limits RLNet's performance on our shorter and noisier TESS LCs.

In summary, CALNet exhibits superior performance due to its specific tailoring to address the unique characteristics of our dataset. It efficiently integrates LC and GLS data while leveraging sophisticated mechanisms like CBAM and LSTM. Its adaptability and robustness render CALNet the top-performing model for our specific task.

\section{Results and discussion}  \label{sec5}
We applied the trained CALNet model to all TESS data from Sectors 1--74. The model demonstrated a high recall rate of 99.1\%, correctly identifying 4,187 out of 4,225 known EBs. Notably, it also led to the discovery of new EBs.
 
\begin{figure}[ht!]
    \centering
    \begin{minipage}[t]{\linewidth}
        \centering
        \includegraphics[width = \linewidth]{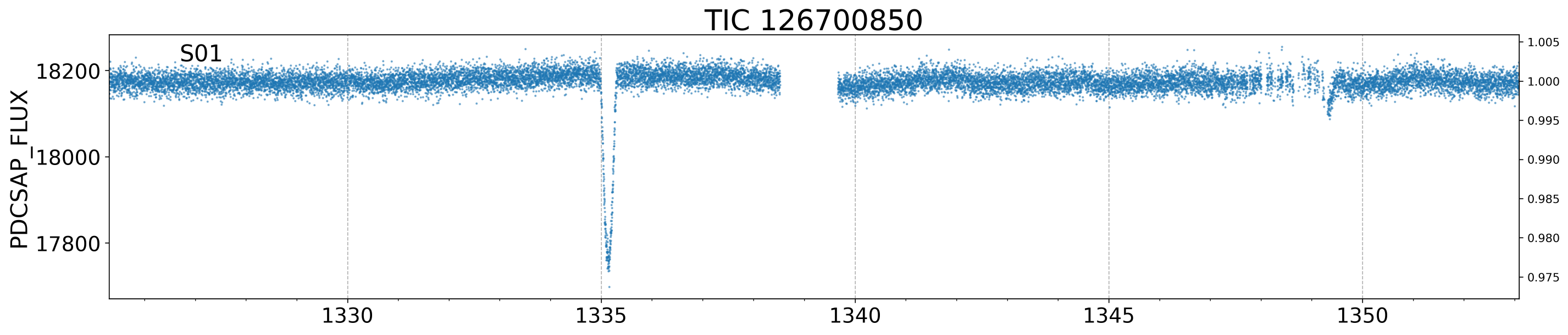}
    \end{minipage}
    \begin{minipage}[t]{\linewidth}
        \centering
        \includegraphics[width=\linewidth]{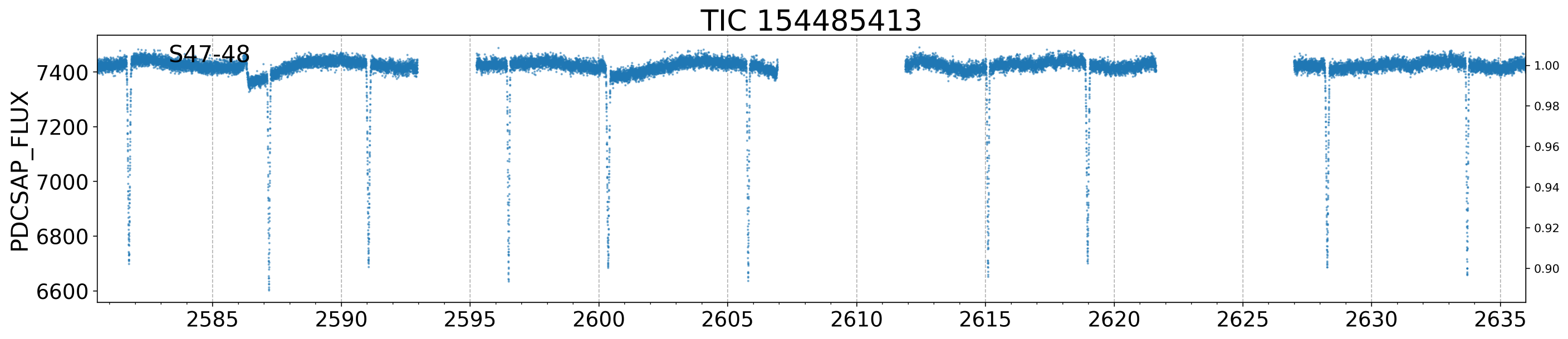}
    \end{minipage}
    \begin{minipage}[t]{\linewidth}
		\centering
		\includegraphics[width=\linewidth]{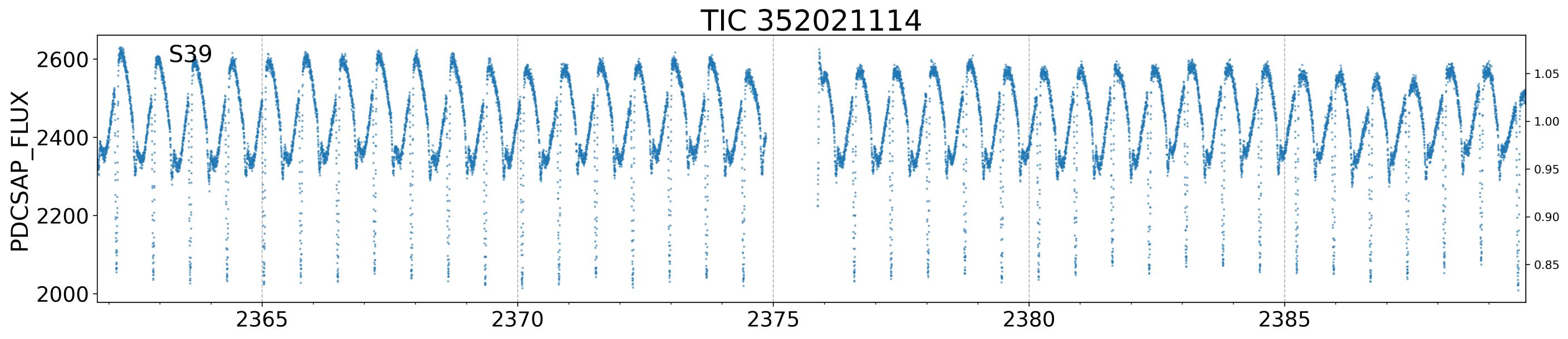}
    \end{minipage}
     \caption{Examples of light curves of three eclipsing binaries identified in this work. X-axes are \texttt{BJD} $-$ 2457000 days, and Y-axis are \texttt{PDCSAP\_FLUX} in TESS 2-minute cadence data products. The ticks on the right sides of each panel show the relative flux normalized to their median values.\label{fig11}}
\end{figure}

After processing with CALNet, each sample received a score between 0 and 1, with scores nearer to 1 suggesting a higher probability of being classified as an EB. To prioritize high recall, we set a threshold of 0.9, meaning that samples with scores $\ge 0.9$ were classified as EBs, while those below the threshold were labeled as non-EBs. It is commonly recognized that prioritizing high recall typically entails sacrificing some precision, prompting us to visually verify the model's predictions. Ultimately, we identified 9351 EBs. Figure \ref{fig11} displays the LCs of several EB candidates, as determined by the model and validated through manual inspection.

\begin{figure}[ht!]
    \centering
    \begin{minipage}[t]{\linewidth}
        \centering
        \includegraphics[width=\linewidth]{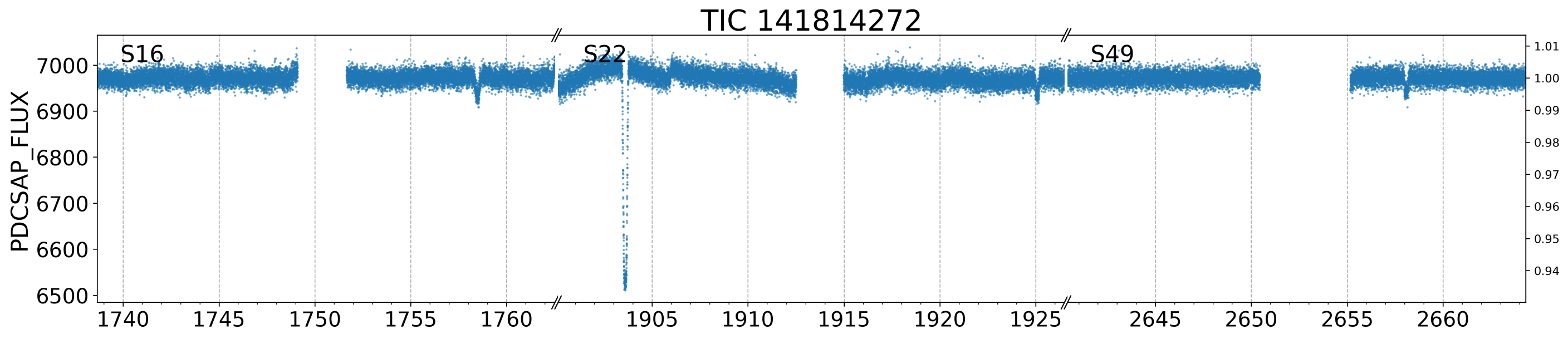}
    \end{minipage}
    \begin{minipage}[t]{\linewidth}
	\centering
	\includegraphics[width=\linewidth]{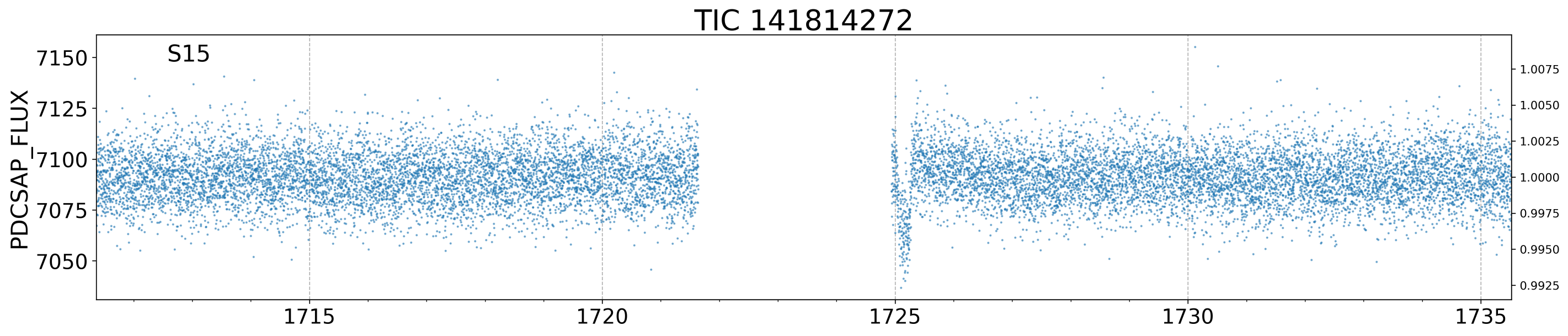}
    \end{minipage}
    \caption{The upper panel shows the light curve of TIC\,141814272, which is judged as a positive detection of eclipse in Sector\,16, 22, and 49, but a negative detection in Sector\,15.}\label{fig12}
\end{figure}

During the manual inspection process, we encountered several challenges, one of which was the unavoidable occurrence of misclassification. Over half of the samples initially predicted as EBs were subsequently found to be incorrect. The LCs of some stars, influenced by their rotation, exhibit patterns that mimic those of EBs, challenging our model's ability to differentiate between them accurately. This issue is tolerable given our primary objective of achieving high recall, which necessarily entails some compromise in prediction accuracy. Furthermore, we noted an intriguing phenomenon where the same star, identified by a unique TIC, could be categorized differently by the model. Upon further analysis, we found that this discrepancy may arise from the observational coverage of the star’s orbit. Specifically, for stars with relatively long orbital periods, eclipses may either be unobservable or insufficiently prominent in certain sectors. In our work, data from different sectors were treated as separate inputs to the model. Consequently, a star may be classified as an EB in sectors where eclipses are observed, but not in those where eclipses are absent. To investigate this hypothesis, we selected a few samples and plotted their LC time series for detailed analysis in Figure \ref{fig12}.

As shown in Figure \ref{fig12}, the LC displays an obvious eclipse in Sector 22, but in Sector 15, the eclipse is not particularly prominent and happens to occur near the edge of blank segments, which may make it challenging for our model to detect. The inconsistency in the observed LCs across different sectors helps explain why the same star, identified by the same TIC, could be classified into different categories by the model.

\cite{Gao2025} identified 72,505 periodic variable stars using TESS two-minute cadence data, among which 5,643 were classified as eclipsing binaries (EA/EB/EW). We cross-matched these eclipsing binaries with our work and found 2,585 common sources in our training sample and 2,386 in our final catalog, leaving 672 sources that were absent in our sample. On the other hand, our catalog has 4,450 common sources with the full sample of \cite{Gao2025}, among which 2,064 were classified as other variable types in their work, with most of them as rotational (1,342) or Cassiopeia variables (593).

\section{Conclusion} \label{sec6}
In this paper, we proposed a deep-learning method for identifying EBs with LC and GLS data. Our main contributions can be summarized as follows:

(1) Proposed a novel LC preprocessing method: We introduced a preprocessing method tailored to the characteristics of our task. We opted against commonly used preprocessing techniques, such as period calculation and phase folding, which often require manual intervention and result in a substantial workload. Given the large volume of data in our task, we designed a simpler preprocessing method using spline interpolation fitting, converting LCs into vectors suitable for model input.

(2) Development of CALNet: We developed CALNet, a hybrid deep learning model integrating CNN, LSTM, and an Attention Block called CBAM, for classifying EBs from TESS data. CALNet achieves a significant improvement in recall, increasing from 97.2\% to 99.1\% compared to our basic CNN model.

(3) Fusion of multidimensional data: By integrating LC features and GLS features, CALNet effectively increased recall from 95.8\% to 99.1\%, compared to traditional models that rely on a single data type. This demonstrates the benefits of utilizing multidimensional data for EB classification. 

(4) Discovery of new EBs: Applying CALNet to TESS data from Sectors 1--74 resulted in a high recall of 99.1\%. Also, we identified 9,351 new EBs following manual visual inspection, expanding the catalog of known samples.

(5) Broad applicability and open-source availability: CALNet is a versatile model applicable to other object identification tasks beyond EBs. The source code has been made publicly available on Github\footnote{\url{https://github.com/wangleon/CALNet}}, fostering further research and extensions by the community.

While our approach demonstrates significant improvements in recall and the identification of new EBs, some challenges remain. For instance, the current model does not provide detailed parameters for the newly identified EBs, which may limit direct application in detailed astrophysical studies. Moreover, classification accuracy could be further validated with more extensive data from future missions.

Despite these limitations, this work highlights the transformative potential of deep learning in astronomy. By integrating advanced preprocessing, multidimensional data fusion, and hybrid architectures, we provide a robust framework for celestial object classification and discovery. This study underscores the importance of leveraging modern AI techniques to process and analyze the vast amounts of data generated by space missions, paving the way for more precise and efficient exploration of the universe.

\begin{acknowledgments}
This work was supported by the National Natural Science Foundation of China (Grant numbers 12473104 and U2031144). The authors are also deeply grateful for the constructive suggestions given by anonymous reviewers.
\end{acknowledgments}

\facilities{TESS}
\software{
    Numpy \citep{numpy},
    Tensorflow \citep{abadi2016tensorflow},
    Keras \citep{keras}
    Astropy \citep{astropy},
    Matplotlib \citep{matplotlib}
}

\bibliography{CALNet}{}
\bibliographystyle{aasjournal}

\end{document}